\documentstyle[12pt]{article}
\newcommand{\be}{\begin{equation}}
\newcommand{\ee}{\end{equation}}
\newcommand{\bea}{\begin{eqnarray}}
\newcommand{\eea}{\end{eqnarray}}

\begin{document}

\centerline{\bf Zero-mode contribution to the light-front Hamiltonian of }
\centerline{\bf Yukawa type models}

\vskip 0.1in
\centerline{A.B. Bylev, V.A.  Franke, E.V. Prokhvatilov}
\centerline{Institute of Physics, St. Petersburg State University}

\begin{abstract}

Light-front Hamiltonian for  Yukawa type models is determined
without the framework of  the canonical light-front formalism.
Special attention is given to the contribution of zero modes.
\end{abstract}

\section{Introduction}

During the last  years, quantization on the light front has
been one of the intensively  developing topics in  field
theory (see review  \cite{Brodsky97} and the references
therein). The most attractive feature of the light front
formalism is the simplicity of definition of physical vacuum
state. It is this feature which instils hope that light-front
Hamiltonian approach might give nonperturbative solution of
relativistic bound state problems in strong interaction.
However for a complete attack on QCD to be feasible many
technical obstacles remain still to be overcome. One  of such
obstacles is known as zero-mode problem
\cite{Brodsky97,Maskawa,Burkardt96} and will be the main topic
of the paper.

In  the canonical  light-front formalism zero modes, i.e.
fields that are constant along  $x^{-} = \frac{1}{\sqrt{2}}
(x^0 - x^3) \; (x^{+}=\frac{1}{\sqrt{2}} (x^0 + x^3) $ --
plays the role of time), are not, generally speaking,
dynamical variables. They have to be determined from the
solution of nonlinear operator constraint equations. Only an
approximate solution of such equations is possible and even in
the case of two dimensional scalar field theories it causes
considerable difficulties \cite{Pinsky}. On the other hand
zero modes are needed for computation of the Poincare
generators. As was noticed in \cite{Burkardt91} the neglecting
of zero modes contribution leads, for example, to breakdown of rotational
invariance in the theory with fermions. As the result
 the operator constraint equations in canonical light-front formalism
must be solved before the correct light-front Hamiltonian
$P_{+}=\frac{1}{\sqrt{2}}(P_0 + P_3)$ can even be written
down. For gauge theories the zero-mode problem becomes much
more complicated.

Various attempts were undertaken  to avoid the solving of
constraint equations and to obtain an effective
light-front theory \cite{FrankeLenz}. Nevertheless, many
aspects of the zero-mode problem are not yet completely under
control and there is still some necessity in developing other
ways to construct the light-front Hamiltonians.

It is our intention  to discuss here a method of construction
of light-front Hamiltonians without the framework of canonical
light-front formalism. The method was proposed in
\cite{Bylev96} and applied there to the  scalar field
theories. We investigate here the case of theories with
fermions. We determine the matrix elements of the Hamiltonian $P_{+}$ in the
same Fock  space that is used in the canonical light-front
formalism  but via Green function equations and special
analysis (without exact calculation) of Feynman diagrams for
Green functions to all orders in perturbation theory. Such
analysis allows to pick out the contribution of zero modes to
the light-front Hamiltonian without direct solving the
constraint equations. We find that zero modes lead to
additional contributions in comparison with naive canonical
light-front Hamiltonian. The structure of such terms  is
discussed in detail.

In section 2 we put forward the method to  construct
the light-front Hamiltonian. We apply it to the case of Yukawa
model and discuss contribution of zero modes to the
Hamiltonian.  Section 3 contains general consideration of
Feynman diagrams. In this respect, a convenient technique is
proposed  to select role of zero modes. We conclude in section
4 with a brief summary of the obtained results.

\section{Hamiltonian}

We consider a system characterized by the Lagrangian
\be
L= \bar{\psi} (\gamma^{\mu} i \partial_{\mu} - M) \psi + \frac
{1}{2} \partial_{\mu} \phi \partial^{\mu} \phi - \frac{m^2}{2}
\phi^{2} + g \bar{\psi} \psi \phi   \label{21}
\ee

The theory (\ref{21}) admits canonical quantization on the $x^0 = 0$
surface which we will assume to be fulfilled, that is
\bea
[\phi(y),\partial_0 \phi (x)] \delta (y^0 - x^0) & = &
\delta^{(4)}(y -x)  , \nonumber \\
\{ \psi_{\alpha}(y),
\psi_{\beta}^{\dagger}(x)\} \delta (y^0 - x^0) & = & \delta
_{\alpha \beta} \delta^{(4)} (y - x)  \label{22}
\eea
and all the other (anti-) commutators are zero.

The fields $\phi(x) \mbox{ and } \psi(x)$ satisfy the equations
\bea
(i \gamma^{\mu} \partial_{\mu} - M + g \phi) \psi & = & 0 , \nonumber \\
(\Box + m^2) \phi - g \bar{\psi} \psi & =& 0 . \label{23}
\eea
These equations of motion (\ref{23}) and the equal-time
commutators then yield usual equations for the quantities
$T(\psi_{\alpha_1}(x_1) \ldots  \bar{\psi}_{\beta_1}(y_1)
\ldots \phi(z_1) \ldots)$, where $T$ stands for the $x^0$ --
chronological ordering operation. We will use them below.

Define the operators $a(\tilde{k}), b(\tilde{k} \lambda)
d(\tilde{k} \lambda)$ via fields on the light front
\bea
a(\tilde{k}) & = & \int d^3 \tilde{x} 2 k_{-} e^{i k x}
\phi(x) \mid_{x^{+}=0} , \ \ \ \ \ k_{-} > 0, \nonumber \\
b(\tilde{k} \lambda) & = & \int d^3 \tilde{x}  e^{i k x} \bar{u}(k \lambda)
  \gamma^{+} \psi(x) \mid_{x^{+}=0} , \ \ \ \ \ k_{-} > 0,
\nonumber \\
d(\tilde{k} \lambda) & = & \int d^3 \tilde{x} \bar{\psi}(x)
\gamma^{+} v(k \lambda) e^{i k x} \mid_{x^{+}=0} , \ \ \ \ \ k_{-} > 0 .
\label{24}
\eea
Here and throughout the paper the following notations are
accepted: $\tilde{x} \equiv (x^{-},x^{\perp})$,$ \tilde{k}
\equiv (k_{-}, k_{\perp}), x^{\perp} \equiv (x^1,x^2),
k_{\perp} \equiv (k_1,k_2)$,$ k_{\pm}=\frac{1}{\sqrt{2}}(k_{0}
\pm k_{3}), kx = k_{+} x^{+} + k_{-} x^{-} + k_1 x^1 + k_2
x^2$. $ u(k \lambda) \mbox{ and } v(k \lambda) $ are free
Dirac spinors with usual normalization conditions: $\bar{u}(\tilde{k}
\lambda) u(\tilde{k} \lambda) = 2 M,\ \ \bar{v}(\tilde{k}\lambda)
v(\tilde{k}\lambda) = - 2 M$.

In the canonical light-front formalism the operators $a(\tilde{k}),
b(\tilde{k} \lambda), d(\tilde{k} \lambda) $ and their
conjugate play the role of annihilation and creation
operators and satisfy the following commutation relations
\bea [a(\tilde{k}),
a^{\dagger}(\tilde{p})] & = & (2 \pi)^3 2 k_{-} \delta^{(3)}
(\tilde{k} -\tilde{p}), \nonumber \\
\{b(\tilde{k} \lambda),
b^{\dagger}(\tilde{p} \mu)\} & = & (2 \pi)^3 2 k_{-}
\delta^{(3)} (\tilde{k} -\tilde{p}) \delta_{\lambda \mu},
\nonumber  \\
\{d(\tilde{k} \lambda), d^{\dagger}(\tilde{p}
\mu)\} & = & (2 \pi)^3 2 k_{-} \delta^{(3)} (\tilde{k}
-\tilde{p}) \delta_{\lambda \mu},
\label{25}
\eea
the other (anti-)commutators are zero. In the canonical light-front
formalism these commutation relations are postulated. In our case they
should, in principle, be proved. It can be done at least in framework
of perturbation theory but here we assume the relations (\ref{25})
without any proof.

Simple kinematical arguments \cite{Nakanishi,Bylev96} show that
the operators (\ref{24})
annihilate the physical vacuum $|0\rangle$. This feature and
the commutation relations (\ref{25}) permit us to construct a
light-front Fock space above the physical vacuum from the
basis vectors like
\be
|k_1 \lambda_1 \ldots q_1 \mu_1 \ldots
t_1 \ldots \rangle = b^{\dagger}(\tilde{k}_1 \lambda_1) \ldots
d^{\dagger}(\tilde{q}_1 \mu_1) \ldots a^{\dagger}(\tilde{t}_1)
\ldots |0 \rangle \label{26}
\ee

Consider a set of wave functions $\langle k_1 \lambda_1 \ldots
q_1 \mu_1 \ldots t_1 \ldots | P \rangle$, where $|P\rangle $
is any eigenstate of the operator $P_{\mu}$, i.e. $P_{\mu}
|P\rangle = p_{\mu} |P \rangle $. Determination of matrix
elements of $P_{\mu}$ in the basis (\ref{26}) is equivalent to
finding Schr\"{o}dinger equation for the wave functions. To obtain these
equations we rewrite the wave function in the form of integral
of Bethe-Salpeter (BS) amplitude  $\langle 0| T(\psi (x)
\ldots \bar{\psi}(y) \ldots \phi (z) \ldots )| P \rangle$
\bea
\lefteqn{ \langle k_1 \lambda_1 \ldots q_1 \mu_1 \ldots t_1
\ldots | P \rangle= }\nonumber \\
&& =\int \prod_{i}\left(d^4
x_i e^{i k_i x_i} \delta(x^{+}_i) \bar{u}_{\alpha'_i }(k_i
\lambda_i) \gamma^{+}_{\alpha'_i \alpha_i}\right)
\prod_{j}\left(d^4 y_j e^{i q_j y_j} \delta(y^{+}_j)
\gamma^{+}_{\beta_j \beta'_j} v_{\beta'_j}(q_j \mu_j)\right)
\times \nonumber \\
&&\times \prod_{l}\left(d^4 z_l 2 t_{l -}
e^{i t_l z_l} \delta(z^{+}_l)\right) \langle 0| T(\psi
 _{\alpha_1}(x_1) \ldots \bar{\psi}_{\beta_1}(y_1) \ldots \phi
 (z_1) \ldots )| P \rangle \label{27}
\eea
Here the symbol $T$
means the chronological ordering operation along $x^0$.
Without $T$-ordering the right-hand side of relation (\ref{27})
is just substitution of definitions (\ref{24}).
The representation (\ref{27})  is possible due to
the fact that difference between product of fields and $T$-product of
fields is translationally invariant quantity but the Fourier transform
 of such quantity is proportional to $\delta(\sum k_{i-} + \sum q_{j-} 
+ \sum t_{l-}) = 0$ as longitudinal momenta
 $k_{i -}, q_{j -}, t_{l -}$ are positive.

Let $\langle 0| T(\psi_{\alpha_1}(k_1) \ldots
\bar{\psi}_{\beta_1}(q_1) \ldots \phi (t_1) \ldots )| P
\rangle$ denote the Fourier transform of the BS amplitude $\langle
0| T(\psi_{\alpha_1}(x_1) \ldots \bar{\psi}_{\beta_1}(y_1)
\ldots \phi (z_1) \ldots )| P \rangle$.  Then  one has
\bea
\lefteqn{ \langle k_1 \lambda_1 \ldots q_1 \mu_1 \ldots t_1
\ldots | P \rangle=} \nonumber \\
&& = \int \prod_{i}(\frac{dk_{i +}}{2 \pi} \bar{u}_{\alpha'_i }(k_i
\lambda_i) \gamma^{+}_{\alpha'_i \alpha_i}) \prod_{j}(\frac{d
q_{j +}}{2 \pi} \gamma^{+}_{\beta_j \beta'_j} v_{\beta'_j}(q_j
\mu_j)) \prod_{l}(\frac{d t_{l +}}{2 \pi} 2 t_{l -} ) \times
\nonumber \\
&&\times \langle 0| T(\psi _{\alpha_1}(k_1)
\ldots \bar{\psi}_{\beta_1}(q_1) \ldots \phi (p_1) \ldots )| P
\rangle \equiv \nonumber \\
&& \equiv \prod_l 2 t_{l -}
\prod_i \bar{u}_{\alpha'_i}(k_i \lambda_i)
\gamma^{+}_{\alpha'_i \alpha_i}\prod_j \gamma^{+}_{\beta_j
\beta_j'} v_{\beta'_j}(q_j \beta'_j) \times \nonumber \\
&&\times  \overline{\langle 0| T(\psi _{\alpha_1}(k_1) \ldots
\bar{\psi}_{\beta_1}(q_1) \ldots \phi (t_1) \ldots )| P
\rangle}, \label{28}
\eea
where we have introduced short
notation for the integral over  plus-component of momenta (the overline).

The equations to be found will be obtained from the relations
\bea
\lefteqn{\langle k_1 \lambda_1 \ldots q_1 \mu_1 \ldots t_1 \ldots
| P _{+}|P\rangle =
p_{+} \langle k_1 \lambda_1 \ldots q_1 \mu_1 \ldots t_1 \ldots
| P \rangle = }\nonumber \\
&& = \prod_l 2 t_{l -} \prod_i \bar{u}_{\alpha'_i}(k_i \lambda_i)
\gamma^{+}_{\alpha'_i \alpha_i}
\prod_j \gamma^{+}_{\beta_j \beta_j'} v_{\beta'_j}(q_j \beta'_j) \times
\nonumber \\
&& \times \overline{(\sum_{i} k_{i +} + \sum_{j} q_{j +} + \sum_{l} t_{l +})
\langle 0| T(\psi _{\alpha_1}(k_1) \ldots \bar{\psi}_{\beta_1}(q_1) \ldots
\phi (t_1)
\ldots )| P \rangle} \label{29}
\eea

Let us consider only one of the items of the sum in the right
hand side of equation (\ref{29}). The other items are
considered analogically. From the equations for $T$-products
of fields we get
\bea
\lefteqn{\gamma^{+} \overline{ k_{1 +}
\langle 0|T(\psi(k_1) \ldots )|P\rangle } = \frac{M^2 + k_{1
\perp}^2}{2 k_{1 -}} \gamma^{+} \overline{\langle 0|
T(\psi(k_1) \ldots )|P\rangle }-} \nonumber \\
&& - g (\frac{M - \gamma^{\perp} k_{1 \perp}}{2 k_{1-}} \gamma^{+} +
\frac{1}{\sqrt{2}} \gamma^0 \gamma^{-} ) \int \frac{d^3
\tilde{l}_1}{(2 \pi)^3} \frac{d^3 \tilde{l}_2}{(2 \pi)^3} (2
\pi)^3 \delta^{(3)} (\tilde{k}_1 - \tilde{l}_1 - \tilde{l}_2)
\times \nonumber \\
&& \times \overline{\langle 0| T(\psi(l_1)
\phi(l_2) \ldots )|P\rangle}, \label{210}
\eea
where dots
denote all the other fields that enter in BS amplitude and are
the same in the left- and right- hand sides of equation
(\ref{210}). A term with $\delta$- function which usually has
place in equations for Green functions disappear due to
positivity of $k_{i -}, q_{j -}$.

The first term in equation (\ref{210}) having substituted in
(\ref{29}) gives $((M^2 + k_{1 \perp}^2)/(2 k_{1 -})) $
$\langle  k_1 \lambda_1 \ldots  |P \rangle$ but the second
term in equation (\ref{210}) must be transformed further to be
presented in the form of linear combination of
wave functions. Indeed, it includes
$\gamma^{-} \overline{\langle 0| T(\psi \ldots )|P \rangle}$
instead of $\gamma^{+} \overline{\langle 0|T(\psi \ldots
)|P\rangle}$ that enters in the wave function representation
(\ref{28}). Besides the region of integration in (\ref{210}) over
$l_{1 -},l_{2 -}$ is spread from $- \infty$ to $ + \infty$ and
includes the points $l_{i -}=0$. If the function  $\overline{\langle
0| T(\psi(l_1) \phi(l_2) \ldots )|P\rangle}$ has a behavior like
$\delta(l_{i -})$ it gives  a singular contribution of zero
modes. Revealing such contribution is the main aim of our
consideration.

If $l_{1 -} \neq 0$ one rewrites term with $\gamma^{-}$ in the form of
terms with $\gamma^{+}$ using again the equations for $T$-products
\bea
\lefteqn{\frac{1}{\sqrt{2}} \gamma^0 \gamma^{-} \overline{\langle 0|
T(\psi (l_1) \ldots )|P \rangle} = \frac{M+ \gamma^{\perp}l_{1 \perp}}
{2 l_{1 -}} \gamma^{+} \overline{\langle 0| T(\psi(l_1) \ldots )
|P\rangle }- }\nonumber \\
&& - \frac{g}{2 l_{1 -}} \int \frac{d^3 \tilde{l}_{11}}{(2 \pi)^3}
\frac{d^3 \tilde{l}_{12}}{(2 \pi)^3} (2 \pi)^3 \delta^{(3)}
(\tilde{l}_1 - \tilde{l}_{11} - \tilde{l}_{12})
\gamma^{+} \overline{\langle 0| T(\psi(l_{11}) \phi(l_{12}) \ldots )
|P\rangle} \nonumber \\
&& \label{211}
\eea
We have not written the terms with $\delta$-functions because
fermion fields marked by dots enter in equation (\ref{29}) in
the form $\bar{\psi} \gamma^{+}$  but $\{\gamma^0 \gamma^{-}
\psi(x), \bar{\psi}(y) \gamma^{+} \}\delta (x^0 - y^0)= 0$. As
a result we obtain
\bea
\lefteqn{\gamma^{+} \overline{ k_{1 +}
\langle 0|T(\psi(k_1) \ldots )|P\rangle } = \frac{M^2 + k_{1
\perp}^2}{2 k_{1 -}} \gamma^{+} \overline{\langle 0|
T(\psi(k_1) \ldots )|P\rangle }-} \nonumber \\
&& - g  \mbox{PV} \int \frac{d^3 \tilde{l}_1}{(2 \pi)^3} \frac{d^3 \tilde{l}_2}
{(2 \pi)^3} (2 \pi)^3 \delta^{(3)} (\tilde{k}_1 - \tilde{l}_1 -
\tilde{l}_2) (\frac{M - \gamma^{\perp} k_{1 \perp}}{2 k_{1-}}
+  \frac{M + \gamma^{\perp} l_{1 \perp}}{2 l_{1 -}} ) \times
\nonumber \\
&& \times \gamma^{+} \overline{\langle 0|
T(\psi(l_1) \phi(l_2) \ldots )|P\rangle} + \nonumber \\
&& + g^2 \mbox{PV} \int \frac{d^3 \tilde{l}_{11}}{(2 \pi)^3} \frac{d^3
\tilde{l}_{12}}{(2 \pi)^3} \frac{d^3 \tilde{l}_2}{(2 \pi)^3}
(2 \pi)^3 \delta^{(3)} (\tilde{k}_1 - \tilde{l}_{11} -
\tilde{l}_{12} - \tilde{l}_2) \frac{\gamma^{+}}{ 2 (l_{11 -} +
l_{12 -})} \times \nonumber \\
&& \times \overline{\langle 0|
T(\psi(l_{11}) \phi(l_{12}) \phi(l_2) \ldots )|P\rangle } +
\ldots \label{212}
\eea
where dots mean possible singular
contribution of the mode $l_{1-}=0$ that comes from the
$ \overline{\langle 0| T(\psi(l_1) \phi(l_2) \ldots
)|P\rangle}$. In respect to $l_{1-}$ the integrals in equation (\ref{212}) 
are taken in
the sense of principal value that is marked by $\mbox{PV} \int$ (i.e.
$l_{1-} =0$ is excluded).

 The first term in (\ref{212}) gives
the following contribution to the Schr\"{o}dinger equation
(\ref{29})
\be
p_{+} \langle k_1 \lambda_1 \ldots  \ldots | P
\rangle = \frac{M^2 + k_{1 \perp}^2}{2 k_{1 -}}  \langle k_1
\lambda_1 \ldots  \ldots | P \rangle + \dots
\ee
Let us discuss the contribution to the Schr\"{o}dinger equation from other
terms of equation (\ref{212}). It proves convenient to define
at this stage the connected Bethe-Salpeter amplitude $\langle
0|T(\psi \ldots \bar{\psi} \ldots \phi \dots)|P\rangle_{c}$ as
an amplitude which cannot be presented in the form like
$\langle 0|T(\ldots)|0 \rangle \langle 0| T(\ldots)|P
\rangle$, where dots denote fields belonging to subsets into
which the set
$\psi,\ldots,\bar{\psi},\ldots,\phi,\ldots$ is divided.
Arbitrary  BS amplitude can be expanded into a sum of connected
BS amplitudes multiplied by connected Green functions. For
example, we can symbolically write for the BS amplitude in the
second term of equation (\ref{212})
\bea
\lefteqn{\overline{\langle 0|T(\psi(l_1) \phi(l_2)
\ldots)|P\rangle} = \overline{\langle 0|T(\psi(l_1) \phi(l_2)
\ldots)|P\rangle}_{c} + }\nonumber \\
&&+ \sum \overline{\langle
0|T(\psi(l_1) \ldots)|0 \rangle}_{c}\ \overline{\langle 0|T(
\phi(l_2) \ldots)|P\rangle}_{c} + \nonumber \\
&& + \sum \overline{\langle 0|T(\phi(l_2) \ldots)|0\rangle}_{c}\
\overline{\langle 0|T(\psi(l_1)  \ldots)|P\rangle}_{c}
\label{213}
\eea
In this decomposition we take into account
that $l_{1 -} + l_{2 -} = k_{1 -} > 0$ and all the other
longitudinal  (minus-component) momenta are positive. The sum
in (\ref{213}) is spread over all possible unordered
partitions of the set of fields
$\psi(k_2),\ldots,\bar{\psi}(q_1),\dots,\phi(t_1),\ldots$ into
subsets and the dots denote fields belonging to these subsets.

Introduction of connected BS amplitude is justified by the
fact that the wavefunctions $\langle k_1 \lambda_1 \ldots q_1
\mu_1 \ldots t_1 \ldots |P\rangle$ have as the integrand in
equation (\ref{28}) only connected BS amplitude due to
positivity of corresponding $k_{-}, q_{-},t_{-}$. Having
substituted the decompositions like (\ref{213}) into equation
(\ref{212}) and then into equation (\ref{29}) one transforms
the right-hand side of (\ref{29})
to desirable form except for the region of values for
variables $l_{i -}$. In wave functions all of the longitudinal
momenta must be positive.

As we will see in next section
the function $\overline{\langle 0|T((\gamma^{+}\psi )\ldots
(\bar{\psi} \gamma^{+}) \ldots \phi \dots)|P\rangle_{c}}$
is
zero if at least one of the longitudinal momenta of the fields is 
negative and
it has no singular contribution of zero modes. Therefore, for
connected BS amplitude the region of integration over longitudinal
momenta in equation (\ref{212}) is, in fact, restricted to
$l_{i -} > 0$. That also means that all zero mode contributions to
Schr\"{o}dinger equation and Hamiltonian can come only
from the factors $\overline{\langle 0|T((\gamma^{+}\psi
)\ldots (\bar{\psi} \gamma^{+}) \ldots \phi \dots)|0
\rangle_{c}}$ in the decomposition into connected components
and from  the function $\overline{\langle 0|T((\gamma^{-}\psi
(l_1)) (\gamma^{+} \psi(k_2))\ldots (\bar{\psi} \gamma^{+})
\ldots \phi \dots)|P\rangle_{c}}$ (see equation (\ref{210})
and below)

In next section we will analyse these  questions via
consideration of Feynman diagrams for connected Green
functions. Let us briefly summarize here some results of this
consideration. First of all, the function $\overline{\langle 0 | T( \ldots
)|0 \rangle}_{c}$, where dots mean fields $\gamma^{+} \psi,
\bar{\psi} \gamma^{+}, \phi,$ with the number of fields more
than two is proportional to $\prod_{i} \delta(k_{i -}),$ where
$k_{i -}$ are longitudinal momenta of the fields.  Secondly,
two-point functions have a term proportional to the $\prod_{i}
\delta(k_{i -})$  and an additional term which is
\be
\gamma^{+} \overline{\langle 0|T(\psi(l)
\bar{\psi}(k))|0\rangle_{c}}\gamma^{+} = (2 \pi)^3
\delta^{(3)} (\tilde{l} +\tilde{k}) \theta(0) \gamma^{+},   \
\ \ k_{-} >0
\ee
for fermion fields and
\be
\overline{\langle0|T(\phi(l) \phi(k))|0\rangle_{c}} = (2
\pi)^3 \delta^{(3)} (\tilde{l} +\tilde{k}) \frac{\theta(0) }{2
k_{-}},   \ \ \ k_{-} >0
\ee
for boson fields. $\theta (x)$ is the
step function.

These additional terms reflect the fact that
nonzero modes of the fields on the light front
satisfy commutation relations (\ref{25}).

Thirdly, $\overline{\langle 0|T((\gamma^{-}\psi (l_1))
(\gamma^{+} \psi(k_2))\ldots (\bar{\psi} \gamma^{+}) \ldots
\phi \dots)|P\rangle_{c}}$ has a singular part (due to
zero mode $l_{1-} =0$) which is
\bea
\lefteqn{\overline{\langle
0|T(\gamma^{-} \psi(l_1)  \gamma^{+} \psi(k_2) \ldots
\bar{\psi}(q)\gamma^{+} \ldots \phi(t)
\ldots)|P\rangle_c}|_{\mbox{zero mode $l_{1-}$}} =} \nonumber \\
&& =  g A(\tilde{l}_1) \left\{ \sum_{i=1}^{n_b} \frac{1}{2 t_{i-}}
\overline{\langle 0|T(\underbrace{\gamma^{-} \psi(l_1)
}\psi(l_1 + t_i) \gamma^{+} \psi(k_2) \ldots
\bar{\psi}(q)\gamma^{+} \ldots \underbrace{\phi(t_i)}
\ldots)|P\rangle_c} + \right. \nonumber \\
&& \left. +  \sum_{i=1}^{n_f}(-1)^{n_f + i}
 \overline{\langle 0|T(\underbrace{\gamma^{-}
\psi(l_1) }\phi(l_1 + q_i) \gamma^{+} \psi(k_2) \ldots
\underbrace{\bar{\psi}(q_i)\gamma^{+}} \ldots \phi(t)
\ldots)|P\rangle_c}\right\} \nonumber \\ 
&&\label{215}
\eea
where $A(\tilde{l}) \sim \delta(l_{-})$ and $A(\tilde{l}) =
\frac{1}{2} \gamma^{-} \gamma^{+} \int \frac{d l_{+}}{2 \pi}
D(l) S^{-1}(l) G(l) \gamma^{-} \gamma^{+}$ and $D(l) = i /
(l^2 - M^2 +i \epsilon)$, $S(l) = i / (\not l - M +
i\epsilon)$, $G(l)$ is the Fourier transform of $\langle
0|T(\psi(x) \bar{\psi}(y))|0\rangle$ ($G(l) = \int d^4 x \exp
(i l (x-y)) \langle0|T(\psi(x) \bar{\psi}(y))|0\rangle$). 
Underbrace under some
term means that this term is dropped.

Taking into account enumerated results the decomposition
(\ref{213}), for example, becomes
\bea
\lefteqn{\overline{\langle0|T(\gamma^{+}\psi(l_1)\phi(l_2)\ldots)| P\rangle} =
\overline{\langle0|T(\ldots)|P\rangle}_{c} + } \nonumber \\
&& + \sum_{i=1}^{n_f}(-1)^{n_f + i}
\overline{\langle0|T(\gamma^{+}\psi(l_1)
\bar{\psi}(q_i)\gamma^{+})|0\rangle_c}\
\overline{\langle0|T(\underbrace{\gamma^{+}\psi(l_1)}\phi(l_2) \ldots
\underbrace{(\bar{\psi}(q_i)\gamma^{+})} \ldots )|P\rangle_c}
+ \nonumber \\
&& + \sum_{i=1}^{n_b}\overline{\langle0|T(\phi(l_2)\phi(t_i))|0\rangle_c}\
\overline{\langle0|T(\gamma^{+}\psi(l_1) \underbrace{\phi(l_2)}\ldots
\underbrace{\phi(t_i)} \ldots )|P\rangle_c} + \nonumber \\
&&  + \overline{\langle0|\phi(l_2)|0\rangle}\
\overline{\langle0|T(\gamma^{+}\psi(l_1) \underbrace{\phi(l_2)}
\ldots)|P\rangle_c}
\label{216}
\eea
We took into account that fermion fields
enter in (\ref{29}) in one of forms $\gamma^{+} \psi, \bar{\psi}
\gamma^{+}$.

Having substituted (\ref{216}) into (\ref{212}) and then into
(\ref{29}) we obtain the contribution of the second term  from
(\ref{212}) to Schr\"{o}dinger equation
\bea
\lefteqn{ p_{+}
\langle k _1 \lambda_1 \ldots |P\rangle = \ldots - g
\sum_{\sigma_1} \int [dl_1][dl_2] (2 \pi)^3 \delta^{(3)}
(\tilde{k}_1 - \tilde{l}_1 - \tilde{l}_2)  \times} \nonumber \\
&& \times  \bar{u}(k_1 \lambda_1) (\frac{M -\gamma^{\perp}
k_{1 \perp} }{2 k_{1 -}} \gamma^{+} + \gamma^{+}\frac{M
-\gamma^{\perp} l_{1 \perp} }{2 l_{1 -}}) u(l_1 \sigma_1 )
\langle \underbrace{k_1 \lambda_1}l_1 \sigma_1 l_2 \dots
|P\rangle - \nonumber \\
&& - g \sum_{i=1}^{n_f} (-1)^{n_f + i} \bar{u}(k_1
\lambda_1) (\frac{M -\gamma^{\perp} k_{1 \perp} }{2 k_{1 -}}
\gamma^{+} - \gamma^{+}\frac{M +\gamma^{\perp} q_{i \perp} }{2
q_{i -}}) v(q_i \mu_i ) \times \nonumber \\
&&  \times \langle
\underbrace{k_1 \lambda_1} (k_1 +q_i)\dots \underbrace{q_i
\mu_i} \ldots  |P\rangle - \nonumber \\
&& - \frac{g}{2}
\sum_{i=1}^{n_b} \sum_{\sigma}\bar{u}(k_1 \lambda_1)
\left(\frac{M -\gamma^{\perp} k_{1 \perp} }{2 k_{1 -}}
\gamma^{+} + \gamma^{+}\frac{M -\gamma^{\perp} (k_{1
\perp}+t_{i\perp}) }{2 (k_{1 -}+ t_{i-})}\right) u((k_1 +
t_i)\sigma ) \times \nonumber \\
&& \times \langle
\underbrace{k_1 \lambda_1}(k_1+t_i) \sigma  \dots
\underbrace{t_i} \ldots |P\rangle - \nonumber \\ && - \frac{2
g \langle \phi \rangle M}{2  k_{1 -}} \langle k_1 \lambda_1
\ldots |P\rangle \label{217}
\eea
To obtain (\ref{217}) we
used the relation
\be \gamma^{+} = \frac{1}{2 l_{-}}
\gamma^{+} \sum_{\sigma} u(l\sigma) \bar{u}(l\sigma)
\gamma^{+}
\ee
Notation $[dl]\equiv \frac{dl_{-}
dl_{\perp}}{(2 \pi)^3 2 l_{-}}, \ l_{-}>0$ was also
introduced.

With the aim of comparison with canonical light-front Hamiltonian it is
convenient to extract from
(\ref{217})  the expression for the Hamiltonian $P_{+}$ in the
operator form. For example, the first term of (\ref{217}) gives 
the following
operator expression
\bea
&&-g \sum_{\lambda \sigma} \int
[dp][dq][dk] (2 \pi)^3 \delta^{(3)}(p-k-q)\times \nonumber \\
&&\times \bar{u}(p \lambda) (\frac{M -\gamma^{\perp} p_{
\perp} }{2 p_{ -}} \gamma^{+} + \gamma^{+}\frac{M
-\gamma^{\perp} q_{ \perp} }{2 q_{ -}}) u(q \sigma )\,
b^{\dagger}(\tilde{p}\lambda) b(\tilde{q}\sigma) a(\tilde{k})
\label{228}
\eea
It is exactly the same term that appears in
naive canonical light-front formalism without zero modes.
Analogically, one can determine operator expressions for
other terms of (\ref{217}) and establish the correspondence with
naive canonical expressions for all terms except  the last
one which represents effective contribution of zero modes and
is equal to
\be
\sum_{\lambda} \int [dk] (-1)\frac{2 g \langle \phi \rangle M}{2
k_{-}} b^{\dagger}(\tilde{k} \lambda) b(\tilde{k} \lambda)
\label{218}
\ee

Similar consideration of the third term in (\ref{212}) leads
to  operator terms some of which coincide with canonical
ones, and we do not discuss them. We  concentrate only on the
difference between our and naive canonical Hamiltonian. This
difference is connected with zero modes.  In the decomposition of the
function
$\overline{\langle 0|T(\psi(l_{11})\phi(l_{12})\phi(l_2) \ldots
|P\rangle}$ into connected components we  have the following 
singular zero mode terms
\be
\overline{\langle 0|T(\phi(l_{12}) \phi(l_2))|0\rangle}_{c}\
\overline{\langle0|T(\psi(l_{11}) \underbrace{\phi(l_{12})
\phi(l_2)} \ldots )|P\rangle_{c}},
\ee
\be
\overline{\langle 0|\phi(l_{12}) |0\rangle} \ \
\overline{\langle 0| \phi(l_2)|0\rangle}\ \
\overline{\langle0|T(\psi(l_{11}) \underbrace{\phi(l_{12})
\phi(l_2)} \ldots )|P\rangle_{c}},
\label{219}
\ee
\be
\overline{\langle 0|\phi(l_{12})
|0\rangle} \ \ \overline{\langle0|T(\psi(l_{11})\underbrace{\phi(l_{12})}
\phi(l_2)\ldots )|P\rangle_{c}} 
\ee
and
\be
 \overline{\langle 0|
\phi(l_2)|0\rangle}\ \  \overline{\langle0|T(\psi(l_{11})
\phi(l_{12}) \underbrace{\phi(l_2)} \ldots )|P\rangle_{c}}
\ee
They generate the
addition to the fermion mass term
 \be
 g^2 \sum_{\sigma} \int
[dk] b^{\dagger}(\tilde{k}\sigma) b(\tilde{k}\sigma) \frac{1}{2 k_{-}}\left (
\langle\phi\rangle^2 + \int \frac{d^4 l_1}{(2 \pi)^4}
\frac{d^4 l_2}{(2 \pi)^4} \frac{2 k_{-}}{2 (k_{-} + l_{1-})}
\langle0|T(\phi(l_1)\phi(l_2))|0\rangle_{c}\right )
\label{229}
 \ee
 and the following interaction that must be added
to (\ref{228})
 \be
 g^2 \langle \phi \rangle \sum_{\lambda
\sigma} \int [dp] [dq] [dk] (2 \pi)^3 \delta^{(3)} (\tilde{p}-
\tilde{q} - \tilde{k}) b^{\dagger}(\tilde{p} \lambda)
b(\tilde{q} \sigma) a(\tilde{k})
\bar{u}(p \lambda) (\frac{\gamma^{+}}{2 p_{-}} +
\frac{\gamma^{+}}{2 q_{-}}) u(q \sigma) \label{56}
 \ee
 Note that the last term in
(\ref{229}) can be presented as a sum of two parts:  first
one is
 $$ \int \frac{d^3 \tilde{l}}{(2 \pi)^3} \frac{2k_{-}}{2
|l_{-}|2 (k_{-} + l_{-})} $$
 and it is called in the literature
 \cite{Pauli85} the self-induced inertia. In canonical
 light-front formalism it arises from normal ordering of the
 seagull terms \cite{Pauli85}. The second part actually
 presents zero mode contribution and is absent in naive
 canonical expression. It is  the $\langle0|(\phi
 -\langle\phi\rangle)^2|0\rangle_0$ ( subscript $0$ indicates
 that we take into account only zero mode).

Now let us discuss zero mode contribution caused by
(\ref{215}). Substituting (\ref{215}) into (\ref{210}) and
then into (\ref{29}) we obtain the following contribution to
Schr\"{o}dinger equation
\bea
\lefteqn{p_{+} \langle
k_1\lambda_1 \ldots |P\rangle = \ldots - \frac{g^2}{2 k_{1-}}
\bar{u}(k_1 \lambda_1) \frac{1}{\sqrt{2}} \gamma^{0} \int
\frac{d^3 \tilde{l}}{(2 \pi)^3} A(\tilde{l}) u(k_1 \lambda_1)
\frac{1}{2 k_{1-}} \langle k_1\lambda_1 \ldots |P\rangle -
}\nonumber \\
&& -  g^2 \sum_{i=1}^{n_b}  \int \frac{d^3
 \tilde{l}_1}{(2 \pi)^3}\int[dl_2] (2 \pi)^3
 \delta^{(3)}(\tilde{k}_1 - \tilde{l}_1 -\tilde{l}_2)
 \bar{u}(k_1 \lambda_1) \frac{1}{\sqrt{2}} \gamma^{0}
 A(\tilde{l}_1) \sum_{\sigma} u((l_1 + t_i) \sigma) \times
 \nonumber \\
&& \times \frac{1}{2(t_{i-} + l_{1-})} \langle
(l_1 + t_i)\sigma  l_2 \underbrace{k_1 \lambda_1} k_2
\lambda_2 \ldots  \underbrace{t_i} \ldots |P\rangle -
\nonumber \\
&& - \sum_{i=1}^{n_f} (-1)^{n_f + i} \int \frac{d^3
\tilde{l}_1}{(2 \pi)^3} \frac{g^2}{2 (l_{1-} + q_{i-})}
\int[dl_2] (2 \pi)^3 \delta^{(3)}(\tilde{k}_1 - \tilde{l}_1
-\tilde{l}_2) \bar{u}(k_1 \lambda_1) \frac{1}{\sqrt{2}}
\gamma^{0} \times \nonumber \\
&& \times A(\tilde{l}_1) v(q_i \mu_i)  
\langle (l_1 + q_i)  l_2 \underbrace{k_1 \lambda_1}
k_2 \lambda_2 \ldots  \underbrace{q_i \mu_i} \ldots |P\rangle
\label{230}
\eea
The first term in (\ref{230}) gives an addition
to the fermion mass term of the light-front Hamiltonian and
agrees with results found by Burkardt \cite{Burkardt91,
Burkardt96}.
\be
\sum_{\lambda} \int[dk] b^{\dagger}(\tilde{k} \lambda) b(\tilde{k} \lambda)
 \frac{-g^2}{2 k_{-}}\left(
\bar{u}(k \lambda) \frac{1}{\sqrt{2}} \gamma^{0}[ \int
\frac{d^3 \tilde{l}}{(2 \pi)^3} A(\tilde{l})] u(k \lambda)
\frac{1}{2 k_{-}}\right) \label{55}
\ee
 The contribution of  other terms of (\ref{230}) to
the Hamiltonian has the following operator form
\bea
&&- g^2 \sum_{\lambda \sigma} \int [dk_1] [dk_2] [dk_3] [dk_4] (2
\pi)^3 \delta^{(3)}(\tilde{k}_1 + \tilde{k}_2 - \tilde{k}_3 -
\tilde{k}_4 )\times \nonumber \\
&& \left( \bar{u}(k_1
\lambda) \frac{1}{\sqrt{2}} \gamma^0 A(\tilde{k}_3 -
\tilde{k}_2)  u(k_3 \sigma) b^{\dagger}(\tilde{k}_1 \lambda)
a^{\dagger}(\tilde{k}_2)
b(\tilde{k}_3 \sigma) a(\tilde{k}_4)\right. + \nonumber \\
&& + \left.\bar{u}(k_1 \lambda) \frac{1}{\sqrt{2}} \gamma^0
A(\tilde{k}_3 - \tilde{k}_2)  v(k_2 \sigma) b^{\dagger}(\tilde{k}_1
\lambda) d^{\dagger}(\tilde{k}_2 \sigma)
a(\tilde{k}_3 ) a(\tilde{k}_4)\right)
\label{331}
\eea
 However, since these operator terms contain
two cteation operators (fermion and boson or antifermion) and,
as consequence, they act on states with at least two corresponding
particles we, to obtain correct expression for the Hamiltonian,
 have to consider the other items in (\ref{29}) which correspond to boson
and antifernions. Analysis analogical to the above-mentioned
 leads for these cases to the  expressions that together with
(\ref{331}) cancel each other.
So the full contribution of such zero mode terms to the Hamiltonian
is zero. For the same reason the zero mode contribution of fermion
and antifermion fields to the boson mass term of the Hamiltonian
 is equal zero.

Thus the full contribution of zero modes to the light-front Hamiltonian
of Yukawa model consists of additions to the fermion mass term (\ref{218}),
(\ref{229}) and (\ref{55}) and additional three-particle interaction
(\ref{56}) and analogical terms with antifermion operators and terms
that are hermitean  conjugate to them.

\setcounter{equation}{0}

\section{Feynman diagram analysis}

To prove statements used  in section 2 let us consider an
arbitrary Feynman diagram that represents connected
$(n+m)$-point connected Green function $W_{n|m} (k_1, \ldots,k_n|
-p_1,\ldots,-p_m)$. Momenta corresponding to amputated lines are
$p_1,\ldots,p_m$, they  are ingoing and $p_{i-} >0$. Momenta
$k_1,\ldots,k_n$ are outgoing. There are no restriction on
the sign of $k_{i-}$.  (For Fourier transform we use the same
definition as in section 2)

The expression for the integrand of the diagram is
\be
I(W)=C(W) \prod_{l=1}^L  P(k_l) \frac{i}{k_l^2 - m_l^2 + i
\epsilon} \prod_{v=1}^V (2 \pi)^4 \delta^{(4)}(P_v -
\sum_{n}\varepsilon_{vn} k_n) \label{31}
\ee
Here $C(W)$
contains all factors belonging to the vertices. $k_l$ are
momenta of both external nonamputated lines $(l=1,\ldots,n)$
and internal lines $(l=n+1,\ldots,L)$, $V$ -- is number of
vertices. We assume at the moment that $V\geq 1$. Special case
with $V=0$ corresponds to free two-point functions. Polinom
$P(k_l)$ is $1$ for scalar lines,  $P(k_l) =
\not\! k_l +m_l$ for fermion lines with one exception that we indicate below.
$P_v =\sum p_{k_{v}}$ is the sum of momenta $p_k$ going into the
vertex $v$. $\varepsilon_{vl}$ is the vertex-edge incidence
matrix of the diagram \cite{Itzykson}. Direction of fermion
lines is chosen along the charge spreading. If the direction of
external fermion line and corresponding momentum are opposite
then for this line $P(k_l)= - \not\! k_l + m_l$.  The orientation
of scalar internal lines is chosen arbitrary.

It proves convenient to use the following integral
representation of free scalar propagator and $\delta$-function
\bea
&& \frac{i}{k_l^2 - m_l^2 + i \epsilon} = \int d\alpha_l
\theta(\alpha_l) e^{i \alpha_l (k_l^2 - m_l^2 + i \epsilon)},
\label{32} \\
&&2 \pi \delta(P_{v+} - \sum \varepsilon_{vl}
k_{l+}) = \int d y^{+}_v e^{- i (P_{v+} - \sum
\varepsilon_{vl} k_{l+}) y^{+}_v} \label{33}
\eea
and to change
$P(k_l) \rightarrow P(i \frac{\partial}{\partial \xi_l})e^{- i
k_l \xi_l}\mid_{\xi_l =0}$.

Integration of $I(W)$ over $k_{l+}, (l=1,\ldots, L)$ then
yields
\bea
\lefteqn{I=\int \prod_{l=1}^L \frac{dk_{l+}}{2
\pi} I(W) = C(W) \prod_{v=1}^V (2 \pi)^3
\delta^{(2)}(P_{v\perp} - \sum_{l} \epsilon_{vl} k_{l\perp})
\times }\nonumber \\
&& \times \int \prod_{v=1}^V
\left(dy^{+}_v \delta (P_{v-} - \sum _{l}\varepsilon_{vl}
k_{l-}) e^{- i y^{+}_v P_{v+}}\right) \int \prod_{l=1}^L
\left( d \alpha_l \theta(\alpha_l) P(i
\frac{\partial}{\partial \xi_l}) \times \right.  \nonumber \\
&& \times \left.\delta(2 \alpha_l k_{l-} + \sum_v y^{+}_v
\varepsilon_{vl} - \xi_l^{+}) e^{- i \alpha_l (k_{l\perp}^2 +
m^2_l - i \epsilon) - i \tilde{k}_l
\tilde{\xi}_l}\right)\mid_{\xi_l =0} \label{34}
\eea
Note that
we integrate not only over internal lines but also over
external lines (see (\ref{28})).

Consider at first conditions that are determined by $\delta$-functions
\be
\prod_{l=1}^L \delta(2 \alpha_l k_{l-} + \sum_v y^{+}_v
\varepsilon_{vl} - \xi_l^{+}) \prod_{v=1}^V  \delta (P_{v-} -
\sum _{l^{\prime}}\varepsilon_{vl^{\prime}}
k_{l^{\prime}-}) \label{35}
\ee
Let us
resolve them in respect to $k_{l-} \mbox{ and } y^{+}_v$.
Introducing a matrix $ C_{v v'}=\sum_{l} \varepsilon_{vl}
\frac{1}{2 \alpha_l} \varepsilon_{v' l}$ we get
\bea
k_{l-}&=& \sum_{v v'}\frac{\varepsilon_{vl}}{2\alpha_l} C^{-1}_{v v'}
(P_{v' -} - \sum_{l'} \frac{\varepsilon_{v' l'}
\xi_{l'}^{+}}{2 \alpha_{l'}}) + \frac{\xi^{+}_l}{2 \alpha_l}
\label{36} \\
y^{+}_v & = & - C^{-1}_{v v'} (P_{v' -} -
\sum_{l'} \frac{\varepsilon_{v' l'} \xi_{l'}^{+}}{2
\alpha_{l'}})
\eea
Let us imagine that all external nonamputated lines have
common additional vertex ( $(V+1)$-th vertex) and denote such new diagram
as $G'$. Then the following representation will take place
\cite{Zavialov}
\be
\sum_{v v'}^V C^{-1}_{v v'}a_v b_{v'} =
\frac{1}{D(\alpha)} \sum_{T_2} (\prod_{l \not \in T_2} 2
\alpha_l)(\sum_{half T_2 } a_v)(\sum_{half T_2} b_v)
\label{37}
\ee
for vectors $\vec{a} \mbox{ and }\vec{b}$
having $ \sum_{v=1}^{V+1} a_v =\sum_{v=1}^{V+1} b_v=0$ and
$$D(\alpha) =\sum_{T_1}(\prod_{l\not\in T_1} 2 \alpha_l) $$
Here $T_1$ is a 1-tree of the graph $G'$, i.e. a connected subgraph
containing all vertices of $G'$ and not having cycles. $T_2$
is a 2-tree of the graph $G'$, i.e. a subgraph of $G'$ containing all
vertices, not having cycles and consisting of exactly two
connected components. Applying (\ref{37}) to (\ref{36}) we
obtain
\be
D(\alpha) k_{l-} = \sum_{T_{1l}}(\prod_{j \not \in
T_{1l}} 2 \alpha_j) (\sum_{half T_{1l}} P_{v-}) -
\sum_{l'=1}^L \xi^{+}_{l'} (-1)^{\sigma_{l l'}} (
\sum_{T^{c}_{1ll'}} (\prod_{j\not \in T^{c}_{1ll'}} 2
\alpha_j)) \label{39}
\ee
here we took $P_{(V+1)-}
=-\sum_{v=1}^V P_{v-}$.  $T_{1l}$ denotes 1-trees $T_1$ which
contain the line $l$. The notation $\sum_{half T_{1l}}$ means that the
sum is done over such 2-trees $T_2$ which become 1-trees after
adding the line $l$, and we choose that half of corresponding
1-tree $T_{1l}$  from which the line comes.
$T^{c}_{1ll'}$ is a 1-tree $T_1$ with a cycle that contains
lines $l$ and $l'$ and that converts into 1-tree
after removing any lines of that cycle. At last,
$\sigma_{ll'}=0$ if the directions of the lines $l$ and $l'$ are
agreed in cycle and
$\sigma_{ll'}=1$ otherwise.

So we get
\bea
\lefteqn{\prod_{l=1}^L \delta(2 \alpha_l k_{l-} + \sum_v
y^{+}_v \varepsilon_{vl} - \xi_l^{+}) \prod_{v=1}^V  \delta
(P_{v-} - \sum _{l}\varepsilon_{vl} k_{l-}) =} \nonumber \\
&& = \prod_{l=1}^L \delta( D(\alpha) k_{l-} - \sum_{T_{1l}}(\prod_{j
\not \in T_{1l}} 2 \alpha_j) (\sum_{half T_{1l}} P_{v-}) +
\sum_{l'=1}^L \xi^{+}_{l'} (-1)^{\sigma_{l l'}} (
\sum_{T^{c}_{1ll'}} (\prod_{j\not \in T^{c}_{1ll'}} 2
\alpha_j)) \times \nonumber \\
&& \times \prod_{v=1}^V
\delta(y_v^{+} + \frac{1}{D(\alpha)} \sum_{T_2}(\prod_{j \not \in
T_2} 2 \alpha_j)(\sum_{half T_{2v}} P_{v^{\prime} -})
- \frac{1}{D(\alpha)} \sum_{l^{\prime}=1}^{L} \xi^{+}_{l^{\prime}}
(\sum_{T_{1l^{\prime}}}(\prod_{j \not \in T_{1l^{\prime}}} 2 \alpha_j)
\varepsilon_{T_{2vl^{\prime}}})
)  \nonumber \\
&&\label{310}
\eea

It follows from (\ref{39}) that for external lines $k_{l-}$
can be only nonnegative when $\alpha_{i} \geq 0$ and $\xi^{+}_i =
0$: indeed, $P_{v-}\  (v=1,\ldots,V)$ are positive and
momenta of external lines are outgoing. As a consequence of (\ref{310})
we obtain $I=0$ if at least one of external momenta is
negative.

We will have a singular contribution of zero mode, i.e. $\delta
(k_{l-})$,  if $\sum_{T_{1l}}(\prod_{j \not \in
T_{1l}} 2 \alpha_j) $ $(\sum_{half T_{1l}} P_{v-})=0$ exactly
for all values of $\alpha$s. It has place for diagrams without
amputated lines ( all $P_{v-}=0$). As we saw in previous
section such cases correspond to some vacuum expectation values in
the light-front Hamiltonian. For  theories with
only scalar fields it exhausts all possible contributions of
zero modes. For  theories with fermion fields there is also
another possibility. If some of $\alpha_i$th are exactly zero,
i.e. there are some $\delta(\alpha_i)$, then we also can get $\delta
(k_{l-})$. It is simply to understand that fact from the following
analogy with electrical circuits. If we identify $2 \alpha_i$
as a resistance of the link $i$, $k_{i-}$ as a current in the
link $i$ and $( - y^{+}_v)$ as a potential of the junction $v$
then the equations that are determined by $\delta$-functions
(\ref{35}) are nothing but Ohm's law for the  links of the
circuit that is presented by the diagram $G'$. The joint $(V+1)$-th
vertex of external lines has zero potential by definition. Now
if $\alpha_i=0$ then both ends of the link have the same
potential and we can reduce this link to the one point without
changing currents in the other links. If as a result of such
reduction we obtain a diagram $G''$ that consists of two parts
connected with each other only through this point and if one
part of two has no external amputated lines, i.e. no external
current goes in it, then currents in all links of this part
will be exactly zero. We will call such part of diagram $G'$
 generelized tadpole and denote it $G'_t$. The reason of
appearence $\delta(\alpha)$ in (\ref{34}) is the following.
In expression (\ref{34}) for fermion fields there is $P(i
\frac{\partial}{\partial \xi_l})$ which contains a term with
$\gamma^{+} i \frac{\partial}{\partial \xi_l^{+}}$, and we have
derivative of $\delta$-function $ i \frac{\partial}{\partial
\xi_l^{+}}\delta(2 \alpha_l k_{l-} + \sum_v y^{+}_v
\varepsilon_{vl} - \xi_l^{+})$ that leads to $\delta(\alpha)$.
To see that let us transform (\ref{34}) in a following way. We
introduce 1 in (\ref{34})   in the form $1=\int (
\prod_{v=1}^Vd \lambda_v \delta (\lambda_v -\sum_l
\varepsilon_{vl} \alpha_l))$. Then we resolve
$$\prod_{l=1}^L
\delta(2 \alpha_l k_{l-} + \sum_v y^{+}_v \varepsilon_{vl} -
\xi_l^{+}) \prod_{v=1}^V  \delta (\lambda_{v} - \sum
_{l}\varepsilon_{vl} \alpha_{l}) $$
in respect to $\alpha_l
\mbox{ and }y^{+}_v$ (as we did above in respect to $k_l \mbox{
and }y^{+}_v$) and get $\alpha_l \mbox{ and }y^{+}_v$ as
functions of $k_{-}, \xi^{+}$ which have the same form as
(\ref{36} -- \ref{39})  with rechange $\alpha_l \leftrightarrow
k_{l-}$,$ \ \lambda_v \leftrightarrow P_{v-}$.  The result of
differentiation over $\xi_j^{+}$ can be rewritten in the form
\bea
&&  \left\{\prod_{l}\delta(2 \alpha_l k_{l-} + \sum_v
y^{+}_v \varepsilon_{vl} - \xi_l^{+}) \prod_{v=1}^V  \delta
(P_{v-} - \sum_{l} \varepsilon_{vl} k_{l-})\right \} \times
\nonumber \\
&&\times \left[\sum_{l=1}^L \frac{\partial
\alpha_l}{\partial \xi_j^{+}} i \frac{\partial}{\partial
\alpha_l} + \sum_{v=1}^V \frac{\partial y^{+}_v}{\partial
 \xi_j^{+}} i \frac{\partial}{\partial y_v^{+}}\right]
 \left\{\prod_l (\theta(\alpha_l) e^{- i \alpha_l
 (k_{l\perp}^2 + m^2_l - i \epsilon)} \prod_{v=1}^V  e^{- i
y^{+}_v P_{v+}}\right\} \nonumber \\
&&\label{311}
\eea
where
\bea
\frac{\partial\alpha_l}{\partial\xi^{+}_i} & = &
(-1)^{\sigma_{li}} \frac{1}{D(k)}
\sum_{T^c_{1li}}(\prod_{j\not\in T^c_{1li}} 2 k_{j-}),
\label{312} \\
\frac{\partial y^{+}_v}{\partial \xi_i^{+}}&=&
\frac{1}{D(k)} \sum_{T_{1i}} \varepsilon_{T_{1i}}
(\prod_{j\not\in T_{1i}} 2 k_{j-}) \label{312a}
\eea
$\varepsilon_{T_{1i}}=\pm 1$: it depends on whether the $i$-th
line goes out or in  the half of $T_{1i}$ which contains the
vertex $v$. The sum in (\ref{312a}) is over such $T_{1i}$ which
have the verteces $v$ and $(V+1)$ in different halfs of $T_{1i}$.
$D(k)= \sum_{T_1} (\prod_{j\not\in T_1} 2 k_{j-})$.

As a result we obtain terms with $\frac{\partial}{\partial
\alpha_l} \theta(\alpha_l) = \delta(\alpha_l)$.

Note that we must here consider only derivatives
$\frac{\partial}{\partial \xi^{+}_i}$ for external lines
because for internal lines we do, in fact, integration over
internal longitudinal momenta.

Let $\alpha_l=0$ leads to generalized tadpole subgraph. Until
$\xi_j^{+} \neq 0$ longitudinal momenta in generalized tadpole
subgraph are of the order $ \xi^{+}$ (see (\ref{39})) and, therefore,
$\sum_{T_1} (\prod_{j \not \in T_1} 2 k_{j-}) = O((\xi^{+})^{C
-1})$, where $C$ is the maximal number of independent cycles made
up of lines belonging to $G'_t$ and the line $l$. At the same time
$$ \sum_{T^c_{1il}}(\prod_{j\not \in T^c_{1il}} 2 k_{j-}) =
\left\{ \begin{array}{ll} O((\xi^{+})^{C -1}) & \mbox{if } i
\in G'_t \\ O((\xi^{+})^{C })     & \mbox{if } i \not\in G'_t
\end{array} \right.  $$
So $\frac{\partial \alpha_l}{\partial
\xi_i^{+}} |_{\xi^{+} =0}$ will be different from zero only if
$i$-th line belongs to the subgraph $G'_t$ and in this case
for external line $\frac{\partial \alpha_l}{\partial
\xi_i^{+}} |_{\xi^{+} =0}= (-1)^{\sigma_{il}}\frac{1}{2 k_l}$ 
as it follows from
(\ref{312}).

The derivative $\frac{\partial}{\partial \xi_i^{+}}$ always
appears in combination with $\gamma^{+}$. Therefore for
functions $\overline{\langle 0 |T((\gamma^{+} \psi) \ldots
(\bar{\psi}\gamma^{+}) \ldots \phi \ldots)|P\rangle_c}$
derivative $\gamma^{+} \frac{\partial}{\partial \xi_i^{+}}$
can not appear. Thus for these functions there are not
additional contributions of zero modes.

We want to find singular contribution of zero mode $l_{1-}$ in
$\overline{\langle 0|T(\gamma^{-} \psi(l_1) \gamma^{+}
\psi(k_2)}$ $\overline{ \ldots \bar{\psi}(q)\gamma^{+} \ldots
\phi(t) \ldots)|P\rangle_c}$ which is contained in equation (\ref{210}).
As was
showed above we must take into consideration only the derivative
$\gamma^{+} \frac{\partial}{\partial \xi_1^{+}}$ and it must
belong to the generalised tadpole subgraph. The tadpole fermion
line containing $l_1$-th line can lean on scalar or
antifermion external lines for which the
$\delta(\alpha_l)$ appears. It is obviously that after taking off
$\delta(\alpha_l)$ in (\ref{34}) the contribution of generalized
tadpole subgraphs are factorized, and we can write
\bea
\lefteqn{\overline{\langle 0|T(\gamma^{-} \psi(l_1) 
\gamma^{+} \psi(k_2) \ldots \bar{\psi}(q)\gamma^{+} \ldots
\phi(t) \ldots)|P\rangle_c}|_{\mbox{zero mode $l_{1-}$}} =}
\nonumber \\
&& = \sum_{i=1}^{n_b} \frac{g}{2 t_{i-}}
A(\tilde{l}_1) \overline{\langle 0|T(\underbrace{\gamma^{-}
\psi(l_1) }\psi(l_1 + t_i) \gamma^{+} \psi(k_2) \ldots
\bar{\psi}(q)\gamma^{+} \ldots \underbrace{\phi(t_i)}
\ldots)|P\rangle_c} + \nonumber \\
&& + g \sum_{i=1}^{n_f}(-1)^{n_f + i}
A(\tilde{l}_1) \overline{\langle 0|T(\underbrace{\gamma^{-}
\psi(l_1) }\phi(l_1 + q_i) \gamma^{+} \psi(k_2) \ldots
\underbrace{\bar{\psi}(q_i)\gamma^{+}} \ldots \phi(t)
\ldots)|P\rangle_c} \nonumber \\
&&\label{315}
\eea
where $A(\tilde{l}) =
\frac{1}{2} \gamma^{-} \gamma^{+} \int \frac{d l_{+}}{2 \pi}
D(l) S^{-1}(l) G(l) \gamma^{-} \gamma^{+}$ and $D(l) = i /
(l^2 - M^2 +i \epsilon)$, $S(l) = i / (\not\! l - M +
i\epsilon)$, $G(l)$ is the Fourier transform of $\langle
0|T(\psi(x) \bar{\psi}(y))|0\rangle$ ($G(l) = \int d^4 x \exp
(i(x-y)l) \langle0|T(\psi(x) \bar{\psi}(y))|0\rangle$). As
was shown above $A(\tilde{l}) \sim \delta(l_{-})$.

\section{Summary}

 We have proposed noncanonical approach
to obtain Schr\"{o}dinger equation and light-front Hamiltonian
in the light-front Fock space. In this method we deal with BS
amplitudes and light-front Hamiltonian is extracted from the
equations for these amplitudes. To do this we have also carried out special
analysis of Feynman diagrams. The advantage of the proposed
method is that we obtain light-front Hamiltonian directly in
normal form in respect to light-front annihilation and
creation operators and quite simply get contribution of zero
modes to this Hamiltonian. The terms caused by zero modes include as a
factor initially unknown vacuum expectation values (VEV) such
as $\langle \phi \rangle$, $\langle \phi^2 \rangle$ and some
integrals of $\langle 0|T(\psi(x) \bar{\psi}(y))|0 \rangle$
and these factors are effective manifestation of zero modes.
As discussed in \cite{Bylev96,Burkardt96} if the eigenvalues
and the eiqenvectors of $P_{\mu}$ are known then one can
calculate these VEVs remaining in the light-front approach
despite unknowledge of exact operator  expressions for zero
modes. Thus, solution of field theoretical models in the
light-front Hamiltonian approach needs a self-consistent
simultaneous determination both of the spectrum of $P_{\mu}$ and
these VEVs.

\vskip 1cm
{\bf Acknowledgements} \\
We would like to thank H.J.Pirner, S.Tsujimaru, S.Paston and F.Saradzhev
for useful discussions. We also acknowledge the hospitality given by
Prof. H.J.Pirner during our stay in the Institute of Theoretical
Physics of Heidelberg University where part of this work was done. We thank
Deutsche Forschungsgemeinschaft for the support of these visits. 


\vskip 1cm

\begin{thebibliography}{99}
\bibliographystyle{unsrt}
\bibitem{Brodsky97} Brodsky S J, Pauli H-C and Pinsky S S
1997  {\em  hep-th}/9705477  

\bibitem{Maskawa} Maskawa T and Yamawaki K  \ 1976  {\it Prog.
Theor. Phys.} {\bf 56} 270 \\
Franke V A, Novozhilov Yu V and
Prokhvatilov E V  \ 1981 \ {\it Lett. Math. Phys.} {\bf 5}
239, 437

\bibitem{Burkardt96} Burkardt M \ 1996 \ {\it  Advances Nucl.
Phys.} {\bf 23} 1

\bibitem{Pinsky} Bender C M, Pinsky S S and van de Sande B \
1993 \ {\it Phys. Rev.} D {\bf 48} 816 \\
Pinsky S S and van
de Sande B \ 1994 \ {\it Phys. Rev.} D {\bf 49} 2001 \\
Heinzl
T, Krusche S, Simb{\"u}rger S and Werner E \ 1992 \ {\it Z. Phys.}
C {\bf 56} 415 \\
Xu X \ 1995 \ {\it J. Phys. G: Nucl. Part.
Phys.} {\bf  21} 1437

\bibitem{Burkardt91} Burkardt M and Langnau A \ 1991 \ {\it
Phys. Rev.} D {\bf 44} 3857

\bibitem{FrankeLenz} Franke V A and Prokhvatilov E V \ 1989 \
{\it Sov. J. Nucl. Phys.} {\bf 49} 688 \\
Lenz F, Thies M,
Levit S and Yazaki K \ 1991 \ {\it Ann. Phys., NY} {\bf 208} 1 \\
 Hornbostel K \ 1992\ {\it Phys. Rev} D {\bf 45} 3781 \\
Burkardt M \ 1993\ {\it Phys. Rev.} D {\bf 47} 4628 \\
Franke
V A and Prokhvatilov E V \ 1996 \ {\it Physics of Atomic Nuclei}
{\bf  59} 2030

\bibitem{Bylev96} Bylev A B \ 1996 \ {\it J. Phys. G: Nucl.
Part. Phys. } {\bf 22} 1553

\bibitem{Nakanishi} Nakanishi N and Yamawaki K \ 1977 \ {\it
Nucl. Phys.} B {\bf 177} 15

\bibitem{Pauli85} Pauli H-c and
Brodsky S J  \ 1985 \ {\it Phys. Rev} D {\bf 32} 1993, 2001

\bibitem{Itzykson} Itzykson C and Zuber J-B  \ 1980 \ {\it
Quantum Field Theory} (New York: McGraw-Hill)

\bibitem{Zavialov} Zavialov O I \ 1979 \ {\it Renormalized
Feynman Diagrams} (Moskwa: Nauka) \end{thebibliography}
\end{document}